\documentclass[final]{aipproc}
\layoutstyle{6x9}
\newcommand{\eqb}{\begin{eqnarray}}
\newcommand{\eqe}{\end{eqnarray}}
\newcommand{\diff}{{\rm d}}
\newcommand{\gammapeak}{\gamma_{\rm p}}
\newcommand{\gammamax}{\gamma_{\rm max}} 
\newcommand{\gammamin}{\gamma_{\rm min}}
\newcommand{\gammawind}{\gamma_{\rm w}}

\newcommand{\mnras}{MNRAS}
\newcommand{\apj}{ApJ}
\newcommand{\aap}{A\&A}

\begin{document}

\title{The high-energy gamma-ray 
light curve of PSR~B1259~$-$63}
\author{J.G. Kirk}{%
address={Max-Planck-Institut f\"ur Kernphysik, D-69029 Heidelberg, Germany}
}
\author{Lewis Ball}{%
address={Australia Telescope National Facility, CSIRO, PO Box 276, Parkes, NSW 2870,
Australia}
}
\author{S. Johnston}{%
address={Australia Telescope National Facility, CSIRO, PO Box 76, 
Epping, NSW 1710, Australia}
}
\begin{abstract}
The high-energy gamma-ray light curve of the binary system
PSR~B1259~$-$63, is computed using the approach 
that successfully 
predicted the spectrum at periastron.  
The 
simultaneous INTEGRAL and H.E.S.S. spectra taken 16 days after periastron
currently 
permit both a model with
dominant radiative losses, high pulsar wind Lorentz factor and 
modest efficiency as well as one with dominant adiabatic losses, a slower wind
and higher efficiency. In this paper we shown how 
the long-term light curve may help to lift this degeneracy. 
\end{abstract}
\maketitle
\section{Introduction}
Observations using the H.E.S.S. array of imaging \v{C}erenkov telescopes 
around and after the periastron
passage in early 2004 detected a strong signal in the TeV range
\cite{schlenkeretal05,aharonianetal05}. The measured spectrum is
in excellent agreement in both slope and absolute normalisation 
with that predicted by a model in which the post-shock pulsar wind
electrons have a simple, single power-law distribution 
\cite{kirkballskjaeraasen99}. Significant
night-to-night fluctuations in the TeV light curve as well as an overall
decrease on the timescale of months were also 
observed by H.E.S.S., possibly
correlated with variations in the unpulsed radio emission
\cite{johnstonetal05}.  
Whereas the short timescale fluctuations, 
especially close to periastron, can plausibly be
attributed to departures from spherical symmetry in the structure of the
pulsar wind or the Be~star wind, the long term light curve 
may help constrain the physics of the emission region.

In this paper, we present preliminary results on the light curves
obtained from an
extended version of the model of Ref.~\citep{kirkballskjaeraasen99}. 
Injection of a double power-law electron spectrum, 
similar to that thought to be
injected into the Crab Nebula by its central pulsar
\cite{gallantetal02} is included, as is 
the transition from radiative to 
adiabatic loss mechanisms as the separation of the stars increases. 
The spectral properties of these models are described by 
\citet{kirkballjohnston05}. 

\section{The model}
The pulsar wind that fuels the Crab Nebula
injects into it relativistic 
electrons and positrons (and possibly ions) whose energy
distribution can be approximated as a double power-law:
$Q(\gamma)\propto\left(\gamma/\gammapeak\right)^{-q_1}$, for 
$\gammamin<\gamma<\gammapeak$ and 
$Q(\gamma)\propto\left(\gamma/\gammapeak\right)^{-q_2}$ for 
$\gammapeak<\gamma<\gammamax$ \citep{lyubarsky03b}.
The high-energy index is determined by the slope of the X-ray
spectrum of the Crab Nebula: $q_2\approx 2.2$, in agreement with theories of
first-order Fermi acceleration at relativistic 
shocks \cite{kirketal00,achterbergetal01}. The low energy index follows from
the slope of the radio to infra-red spectrum: $q_1\approx1.6$.  With these
values, most particles are concentrated around the lower cut-off 
at $\gamma=\gammamin$, whereas most of the energy is injected in electrons of
Lorentz factor $\gamma\sim\gammapeak$. In the Crab, $\gammamin\approx100$,
$\gammapeak\approx10^6$ and $\gammamax\approx 10^9$. The resulting synchrotron
spectrum contains two breaks, one due to cooling and one intrinsic to the
injected spectrum (at $10^{13}\,$Hz and $10^{15}\,$Hz in the Crab), 
as well as upper and lower cut-offs. If this injection spectrum is produced at
the termination shock front, and if the cold upstream flow is dominated
by the kinetic energy flux in electron-positron pairs, 
then the Lorentz factor of the wind is 
$\gammawind=\int\diff\gamma\, \gamma Q(\gamma)/\int\diff\gamma\, Q(\gamma)$.
In the following we adopt this injection model.

In PSR~B1259~$-$63, 
relativistic electrons and positrons in the shocked pulsar wind
suffer adiabatic losses as the plasma expands away from the shock
front, as well as radiative losses by synchrotron radiation and inverse Compton
scatterings, primarily of the ultra-violet photons from the Be~star. 
These loss processes have different 
energy dependences, which lead to differences in the 
resulting distribution function. 
The emitted radiation is a combination of synchrotron radiation in a uniform
magnetic field and inverse Compton scattering of ultra-violet photons from the
Be~star. On its way from the pulsar system to the observer the inverse-Compton
emission is partially reabsorbed by pair production on the stellar photons
\citep{kirkballskjaeraasen99}. 
Two sets of models were  constructed in 
Ref.~\citep{kirkballskjaeraasen99}: one for purely
adiabatic and one for purely radiative losses. Both 
were calibrated using the observed X-ray synchrotron emission, and provided 
accurate
predictions of the TeV spectrum subsequently detected 
just before periastron. However, the two models imply quite different injection
spectra. 

As the pulsar moves away from the Be~star, both the target radiation field and
the magnetic field where the winds interact decrease, along with the gas
pressure. For the toroidal field structure 
expected in a pulsar wind, 
the synchrotron loss rate scales with the inverse square of the separation of
the stars. Thus, 
the ratio of the energy densities of 
magnetic field and target radiation remain constant, so that, in the 
absence of Klein-Nishina effects, the ratio of synchrotron to inverse Compton
radiation should not vary with binary phase. 
However, if the expansion time
scales linearly with the separation of the stars, adiabatic losses become
more important with respect to radiative losses as the stars move apart.
In the models used here and in Ref~\citep{kirkballjohnston05}, 
we account for this in the kinetic equation describing the electron
distribution 
by switching between a radiative 
and an adiabatic loss term at the Lorentz factor where the loss
rates coincide. The losses themselves are fixed as functions of binary phase, 
once the magnetic field
strength in the emission region, and the adiabatic loss time scale are
given at periastron. 

\begin{table}
\caption{%
\label{parameters}%
The model parameters. The efficiency refers to the fraction of the
  spin-down luminosity injected into the source as relativistic particles
  (assuming a source distance of $1.5\,$kpc). The adiabatic loss time 
$t_{\rm ad}$ is given in units of the light crossing time of the periastron
  separation ($320\,$sec). $B$ is the 
magnetic field strength in the source at periastron %
}
\begin{tabular}{l|lllllll}
Model:&$\gammamin$&
$\gammapeak$&$\gammamax$&$\gammawind$&$B$&Efficiency&$t_{\rm ad}$\\
\hline         
&&\cr
A1 &$425$ &$10^7$  &  $5\times10^7$ &$5.5\times10^4$ &$0.3\,$G &$10$\% &15  \\
A2 &$425$ &$10^7$  &  $5\times10^7$ &$5.5\times10^4$ &$0.3\,$G &$10$\% &30  \\
B &$425$ &$10^6$ &$4\times10^7$ &$2.9\times10^4$  &$0.3\,$G &$100$\% & 0.5\\
\end{tabular}
\end{table}

\section{Results}
Modelling the spectrum and light curve of the high-energy emission during
the 2004 periastron passage is made difficult by the scarcity of simultaneous 
TeV and X-ray data sets. The only ones currently available
are the X-ray/soft gamma-ray spectrum detected by INTEGRAL 
between 14 and 17 days after
periastron passage \cite{shawetal04} and the March
2004 observations by H.E.S.S. \cite{aharonianetal05}.
\citet{kirkballjohnston05} showed that 
these data are not sufficient to determine the dominant loss mechanism
and discussed two examples. In their first (Model~A) the energy losses suffered
by those particles that emit TeV radiation at around periastron are dominated
by the radiation processes: inverse Compton scattering and
synchrotron radiation. In their second (Model~B) it is assumed that the 
relativistic electrons rapidly move out of the zone where the radiation
is emitted. The spectral fit is in each case good.

\begin{figure}
\hbox{%
\includegraphics[width=0.5\textwidth]{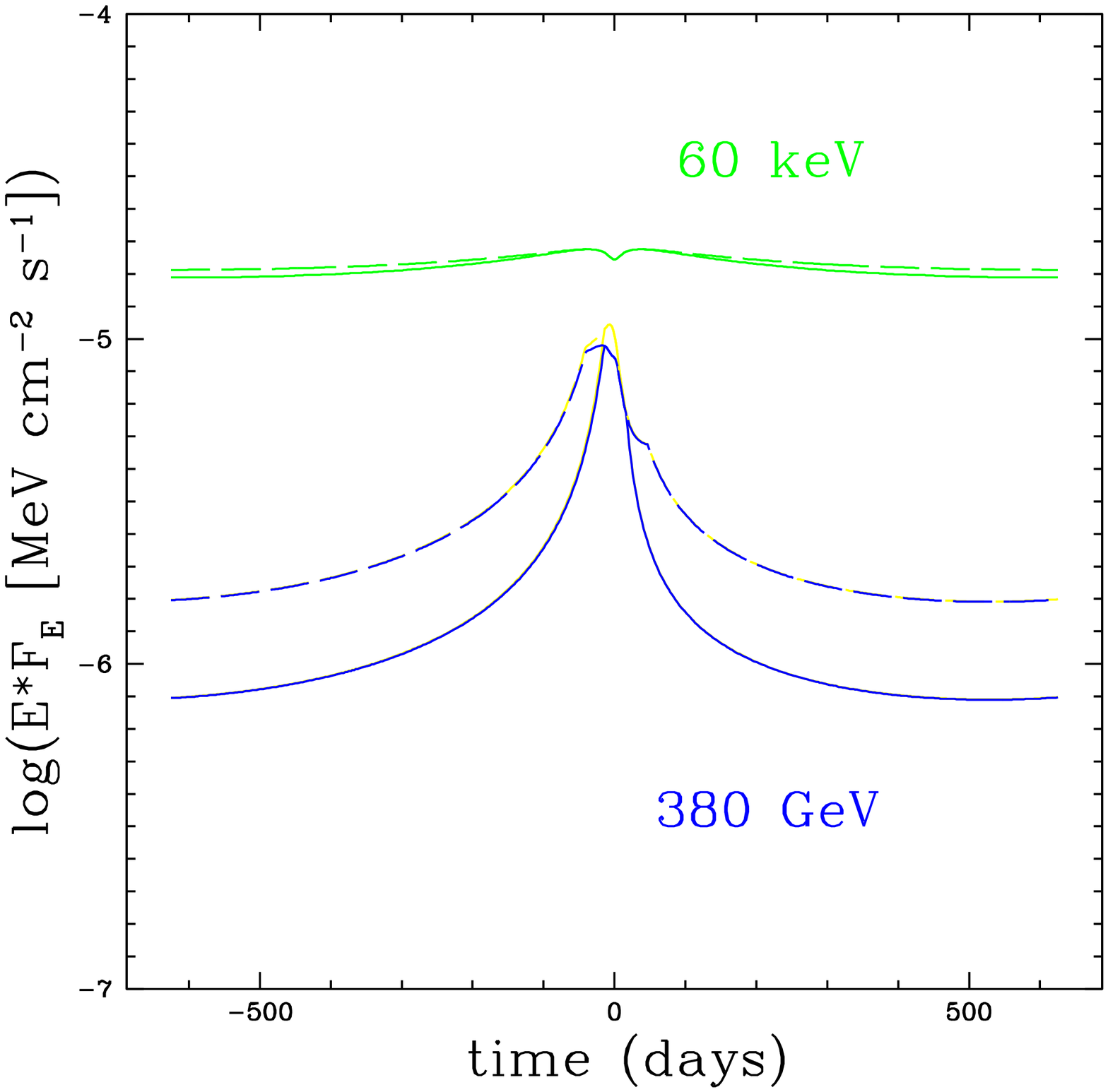}%  
\includegraphics[width=0.5\textwidth]{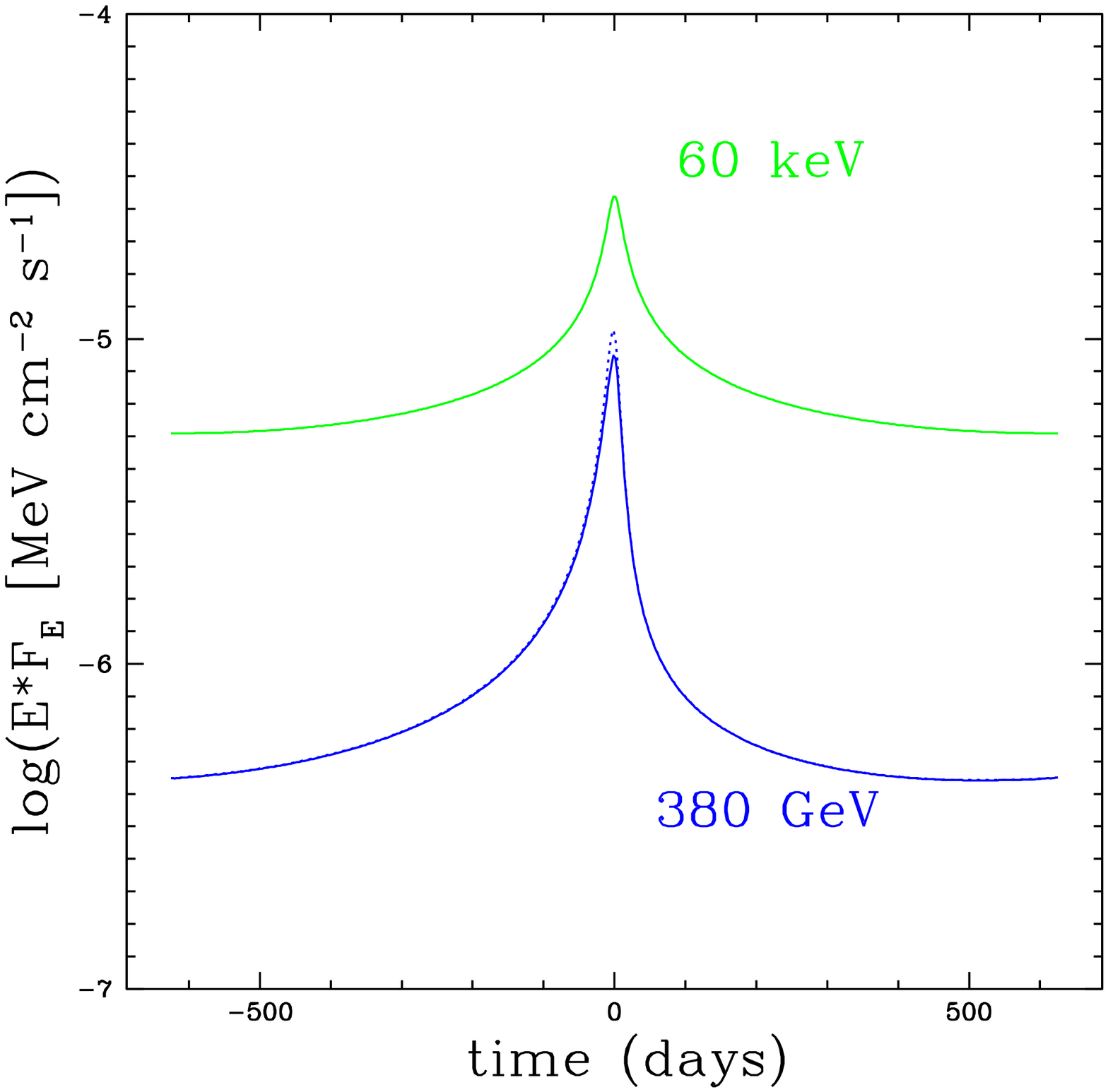}%  
}
\caption{%
\label{lcurve}%
Light curves covering the full binary period of 1236.8 days, centred on
periastron at $t=0$. In Models~A1 and A2 (left-hand panel, solid and 
dashed lines, respectively) 
the losses at periastron of particles
emitting TeV photons are radiative, in Model~ B (right-hand panel)
they are adiabatic.}
\end{figure}

In Fig.~\ref{lcurve} we present results for the light curves over the entire
binary period for three models, whose parameters are given in
Table~\ref{parameters}. 
Model~A1 and Model~B correspond exactly to the two models discussed in 
Ref.~\citep{kirkballjohnston05}. Model~A2 differs from A1 by having double the
adiabatic loss time-scale. 
This makes little difference to the 
hard X-ray and TeV spectra around periastron. However, it changes the 
long term light curve considerably. 

Provided
radiative losses determine the electron spectrum, the synchrotron light curve
remains almost constant. This is because the emission region acts as a
calorimeter; even though the loss rate decreases away from periastron, the
energy radiated remains almost the same. 
This behaviour is evident in the $60\,$keV light curves of Models A1 and A2,
shown in the left hand panel of Fig.~\ref{lcurve}. 
In the case of $380\,$GeV emission, the
time-dependence in the absence of adiabatic losses 
arises from the angle dependence of the
scattering cross section. Because the target photons form an almost
mono-directional beam, the scattering angle needed to deflect a photon into the
direction of the observer is a function of binary phase. As discussed by 
\citet{kirkballskjaeraasen99}, this
leads to a characteristic time dependence that is asymmetric about
periastron. However, the amplitude of the variation is at most a factor of two.

On the other hand, when adiabatic losses dominate, the energy 
radiated by 
an electron in traversing the 
emission region is proportional to the ratio of the 
radiative to the adiabatic loss rates.  
Here we assume that 
the adiabatic loss time scales linearly with the separation of the stars and
the radiative loss time scales quadratically. This implies that
adiabatic losses become relatively more important as the stars move away from
each other. As a result, the light curve shows a strong peak at periastron,
as can be seen in the right-hand panel of Fig.~\ref{lcurve}.
 The ratio of maximum to minimum 
flux density in this model is roughly a factor of 25, corresponding
approximately to the ratio of the apastron to the periastron separation
($14.4$) multiplied by an angle dependent factor $\sim2$.
The $380\,$GeV light curves 
in the left-hand panel show the effects of increasing the adiabatic loss
time-scale by a factor of two. The emission in Model~A1 makes the transition
from radiative to adiabatic losses already at 17 days after periastron. As a
result, the intensity falls off sharply, and the long-term light curve is not
dissimilar to that of Model~B. On the other hand, Model~A2, in which the
adiabatic loss time-scale is twice as long, remains dominated by radiative
losses until roughly 50 days after periastron. (This transition 
is unphysically sharp in our model.) The resulting
ratio of maximum to minimum 
flux density in this model is only a factor of six. 
Although we do not present here a detailed 
comparison with the data, it is clear that the H.E.S.S. light curve 
\citep{schlenkeretal05,aharonianetal05}
favours Models~B and A1 over Model A2.

%\bibliographystyle{aipproc}
%\bibliography{/home/kirk/tex/conferences/Torun/1259}

\end{document}